\documentclass[]{mosi}

\usepackage{helvet}

\usepackage{amsmath} 
\usepackage{natbib}
\usepackage{graphicx}
\usepackage{subcaption} 

\usepackage[toc,page,header]{appendix}
\usepackage[utf8]{inputenc} 
\usepackage[T1]{fontenc}    
\usepackage{hyperref}       
\usepackage{url}            
\usepackage{booktabs}       
\usepackage{lmodern}        
\usepackage{amsfonts}       
\usepackage{nicefrac}       
\usepackage{microtype}      
\usepackage{wrapfig}

\usepackage{amssymb}        
\usepackage{titletoc}
\usepackage{minitoc}

\usepackage{array}
\usepackage{etoolbox}

\definecolor{lightblue}{RGB}{200, 230, 255}  
\definecolor{headerblue}{RGB}{150, 200, 255} 

\usepackage{pgfplots}
\usepackage{xcolor}
\usepackage{float}
\usepackage{comment}
\usepackage{multirow}
\usepackage{makecell}
\usepackage{siunitx}
\usepackage{tikz}
\usepackage{pgf-pie}
\usepackage[export]{adjustbox}

\usepackage{ragged2e}
\usepackage{tabularx}
\usepackage{caption}
\usepackage{enumitem}
\usepackage{pifont}
\usepackage[hang,flushmargin]{footmisc}

\usepackage{tcolorbox}
\tcbuselibrary{breakable}
\tcbuselibrary{skins}

\usepackage{listings}

\usepackage{threeparttable}
\usepackage{setspace}

\usepackage[scheme=plain]{ctex}

\usepackage{algorithm}
\usepackage{algpseudocode}
\usepackage{fontawesome5}
\usepackage{svg} 

\newcommand{\cjk}[1]{#1}

\definecolor{MossCyan}{HTML}{82D9FF} 
\definecolor{MossBlue}{HTML}{82B1FF}

\definecolor{ForestGreen}{RGB}{34, 139, 34}
\definecolor{Red}{RGB}{255, 0, 0}

\definecolor{tickG}{rgb}{0.1, 0.588, 0.1}
\definecolor{crossR}{rgb}{0.588, 0.1, 0.1}

\definecolor{frenchblue}{rgb}{0.0, 0.45, 0.73}
\definecolor{babyblue}{rgb}{0.54, 0.81, 0.94}
\definecolor{classicrose}{rgb}{0.98, 0.8, 0.91}
\definecolor{beige}{rgb}{0.96, 0.96, 0.86}
\definecolor{forestgreen}{HTML}{2e7d43}

\definecolor{blue1}{HTML}{91BBE6}
\definecolor{blue2}{HTML}{3F90E0}
\definecolor{blue3}{HTML}{316FAD}

\definecolor{color1}{HTML}{FF9999}
\definecolor{color2}{HTML}{FF6666}
\definecolor{color3}{HTML}{FF3333}
\definecolor{color4}{HTML}{E60000}
\definecolor{color5}{HTML}{B30000}
\definecolor{color6}{HTML}{8CD98C}
\definecolor{color7}{HTML}{53c653}
\definecolor{color8}{HTML}{00B050}
\definecolor{color9}{HTML}{2d862d}
\definecolor{color10}{HTML}{206020}
\definecolor{color11}{HTML}{cca300}

\newtcolorbox{promptbox}[2][]{
    colback=white,
    coltext=black,
    arc=3mm,
    boxrule=0.5pt,
    colframe=black!60!white,
    title={#2},
    colbacktitle=black,
    coltitle=white,
    fonttitle=\bfseries,
    top=8pt,
    bottom=8pt,
    left=10pt,
    right=10pt,
    breakable,
    before upper={%
        \linespread{1}\selectfont
        \setlength{\parskip}{1ex plus 0.2ex minus 0.2ex}%
        \setlength{\parindent}{0pt}%
    },
    #1
}

\title{MOSS-TTSD: Text to Spoken Dialogue Generation}
\author{SII-OpenMOSS Team\textsuperscript{*}}

\newcommand{\hflogo}{\includegraphics[height=2.0ex]{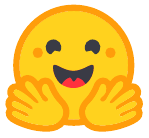}}

\abstract{
\microtypesetup{protrusion=false}
\begin{abstract}

Spoken dialogue generation is crucial for applications like podcasts, dynamic commentary, and entertainment content, but poses significant challenges compared to single-utterance text-to-speech (TTS).
Key requirements include accurate turn-taking, cross-turn acoustic consistency, and long-form stability, which current models often fail to address due to a lack of dialogue context modeling.
To bridge this gap, we present MOSS-TTSD, a spoken dialogue synthesis model designed for expressive, multi-party conversational speech across multiple languages.
With enhanced long-context modeling, MOSS-TTSD generates long-form spoken conversations from dialogue scripts with explicit speaker tags, supporting up to 60 minutes of single-pass synthesis, multi-party dialogue with up to 5 speakers, and zero-shot voice cloning from a short reference audio clip.
The model supports various mainstream languages, including English and Chinese, and is adapted to several long-form scenarios.
Additionally, to address limitations of existing evaluation methods, we propose TTSD-eval, an objective evaluation framework based on forced alignment that measures speaker attribution accuracy and speaker similarity without relying on speaker diarization tools.
Both objective and subjective evaluation results show that MOSS-TTSD surpasses strong open-source and proprietary baselines in dialogue synthesis.

\microtypesetup{protrusion=true}
\end{abstract}
}

\checkdata[\raisebox{-0.1ex}{\faGithub}\hspace{0.4em}Code]{\url{https://github.com/OpenMOSS/MOSS-TTSD}}
\checkdata[\raisebox{-0.1ex}{\hflogo}\hspace{0.4em}Model]{\url{https://huggingface.co/OpenMOSS-Team/MOSS-TTSD-v1.0}}
\checkdata[\raisebox{-0.1ex}{\faChartBar}\hspace{0.4em}TTSD-eval]{\url{https://github.com/OpenMOSS/TTSD-eval}}

\begin{document}
\maketitle

\begingroup
\renewcommand{\thefootnote}{\fnsymbol{footnote}}
\setcounter{footnote}{1}
\footnotetext{Full contributors can be found in the Contributors section.}
\endgroup

\section{Introduction}

Generating spoken dialogue is more challenging than monologue speech due to the need for accurate and natural turn-taking, robust speaker switching, and long-range contextual coherence across multiple turns.
Models must maintain speaker-specific timbre and prosody within and across turns and produce long-form audio without stitching artifacts.
While current text-to-speech synthesis systems (TTS) has made substantial progress for short, single-speaker utterances~\citep{valle, F5-TTS, seed_tts, SparkTTS, Llasa}, gaps remain in coherence, spontaneity, and robustness for real-world conversational scenarios.

To address these challenges, we introduce MOSS-TTSD (Text-to-Spoken Dialogue), a spoken dialogue synthesis model built for expressive, multi-party conversations.
Since its initial release, MOSS-TTSD has evolved through multiple iterations from v0 to v1.0; this paper focuses on the latest release, \textbf{MOSS-TTSD v1.0}.
Unless otherwise specified, all descriptions and evaluations in this paper refer to MOSS-TTSD v1.0.

MOSS-TTSD follows a fully discrete speech generation paradigm: it uses Qwen3-8B-base~\citep{Qwen3} as the autoregressive backbone together with MOSS-Audio-Tokenizer~\citep{gong2026mossaudiotokenizerscalingaudiotokenizers}.
Inspired by MusicGen, we adopt a multi-head delay pattern~\citep{MusicGen} to autoregressively predict RVQ codebook tokens, and we model only the first 16 RVQ layers to enable robust long-context generation at a low bitrate.
Conditioned on dialogue scripts with explicit speaker tags (e.g., \texttt{[S1]/[S2]}) and optional per-speaker reference audio, MOSS-TTSD supports accurate turn-taking and zero-shot multi-speaker voice cloning, and scales to long-form synthesis, generating coherent conversations of up to 60 minutes in a single pass without stitching artifacts.

The key features of MOSS-TTSD can be summarized as:
\begin{itemize}[noitemsep,topsep=0pt]
  \item \textbf{Long-form script-to-conversation generation.} MOSS-TTSD converts dialogue scripts into spoken conversations and supports up to 60 minutes of single-pass generation.
  \item \textbf{Multi-party voice cloning.} MOSS-TTSD supports up to 5 speakers in a single dialogue session and provides zero-shot voice cloning from a short reference audio clip to improve speaker consistency.
  \item \textbf{Multilingual and scenario coverage.} MOSS-TTSD provides strong support for multiple languages, such as English, Chinese, Spanish, Portuguese, German, French, Japanese, Korean, and Russian. It is adapted to scenarios such as podcasts, dynamic commentary, and entertainment content like audiobooks, dubbing, and crosstalk.
\end{itemize}

Additionally, to address the limitations of existing evaluation metrics that rely on speaker diarization tools, we introduce TTSD-eval, an objective evaluation framework that leverages forced alignment to measure speaker attribution, speaker similarity, and intelligibility in long-form, multi-speaker scenarios.
Comprehensive evaluations with TTSD-eval, together with human subjective assessments, demonstrate that MOSS-TTSD outperforms strong open-source and proprietary baselines for spoken dialogue synthesis.

\section{MOSS-TTSD History}

Since June 2025, MOSS-TTSD has evolved through several iterations, progressively improving long-form stability, speaker consistency, and feature coverage.

\textbf{MOSS-TTSD v0 (2025-06).}
We released MOSS-TTSD v0 with bilingual support for Chinese and English, and up to 960\,s single-pass generation.
In the data pipeline, we used WhisperD~\citep{darefsky2024parakeet} for English ASR, and an internal diarization-enabled Whisper model fine-tuned from Whisper-large-v3~\citep{radford2022robustspeechrecognitionlargescale} for Chinese.
Due to the ASR model's 30\,s per-request audio constraint, multi-speaker transcripts were constructed by combining coarse-grained speaker segments from a diarization model with fine-grained, token-level speaker labels predicted by the ASR model, enabling explicit speaker annotation over long recordings.

\textbf{MOSS-TTSD v0.5 (2025-07).}
In v0.5, we improved timbre switching and zero-shot voice cloning robustness by fine-tuning on a high-quality subset of training data curated with a TTSD-eval-like pipeline.
We also introduced a 32\,kHz variant of the XY-Tokenizer to support higher-fidelity synthesis.

\textbf{MOSS-TTSD v0.7 (2025-11).}
v0.7 substantially overhauled the data pipeline by adopting an early internal version of MOSS Transcribe Diarize~\citep{ai2026mosstranscribediarizetechnical} to generate ASR transcripts with explicit speaker tags end-to-end, reducing errors caused by inconsistencies between diarization outputs and transcript-level speaker labeling.
In addition, we refined the XY-Tokenizer decoder so that decoding is explicitly conditioned on both the LLM-generated tokens and the speaker reference tokens, which further improved voice cloning capability and audio quality.
These improvements also extended the maximum single-pass generation length from 960\,s to 1700\,s.

\textbf{MOSS-TTSD v1.0 (2026-02).}
MOSS-TTSD v1.0 further strengthens long-form synthesis, benefiting from scaled data and a stronger LLM backbone, as well as continued improvements in the audio tokenizer and data pipeline.
It supports up to 60-minute single-session context and multi-party interactions with up to five speakers, and expands beyond Chinese and English to multiple additional mainstream languages.
\section{Data Engineering}

\begin{figure}[t]
  \centering
  \includegraphics[width=0.8485\linewidth]{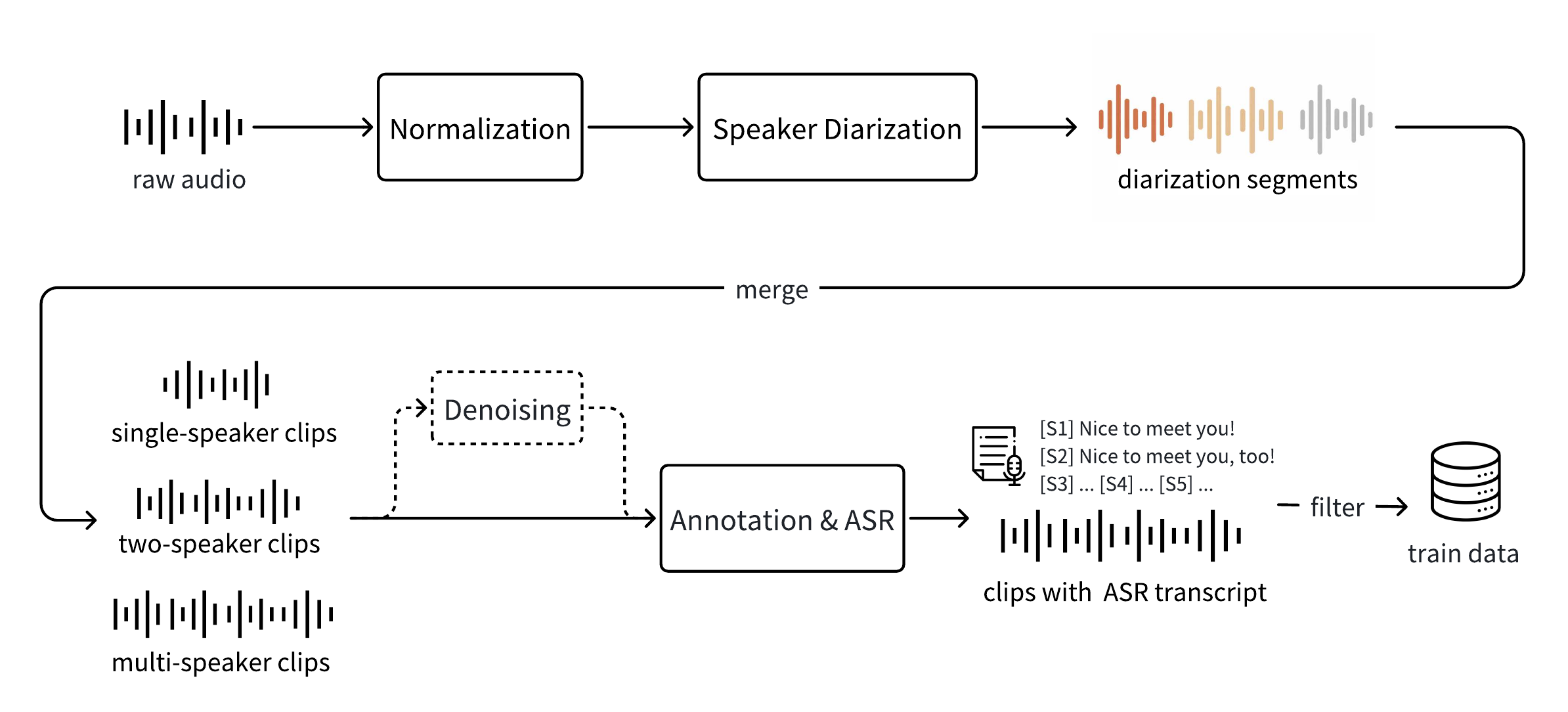}
  \caption{Overview of the MOSS-TTSD data pipeline: raw audio is normalized and diarized, then merged into clips with varying speaker counts and annotated with quality, language, and sampling-rate metadata; noise-heavy domains may undergo additional denoising. Clips are transcribed end-to-end with explicit speaker tags, followed by heuristic and audio/text language-consistency filtering to produce the final training set.}
  \label{fig:data_pipeline}
\end{figure}

\subsection{Data Pipeline}

Figure~\ref{fig:data_pipeline} provides an overview of our data pipeline. MOSS-TTSD shares the same foundational data processing pipeline as MOSS-TTS~\citep{gong2026mossttstechnicalreport}, encompassing raw data collection, audio normalization, and speaker diarization.
Following this baseline processing, we obtain diarization segments. Continuous segments derived from the same raw recording are merged into single-speaker, two-speaker, or multi-speaker (3--5 speakers) clips, with a maximum duration capped at 3600\,s.
Each clip is annotated with a DNSMOS score~\citep{reddy2021dnsmosnonintrusiveperceptualobjective}, a language label generated by Whisper-large-v3~\citep{radford2022robustspeechrecognitionlargescale}, and the true sample rate estimated from the Mel-spectrogram, serving as criteria for subsequent filtering.

Subsequently, leveraging its superior performance in long-context and multi-speaker scenarios, we employ MOSS Transcribe Diarize~\citep{ai2026mosstranscribediarizetechnical} as the ASR model to process each clip in an end-to-end manner, simultaneously yielding the ASR transcript and explicit speaker tags.
We implement a set of heuristic rules to filter out low-quality audio and samples exhibiting ASR hallucinations.
Additionally, a lightweight language detection tool, fastText~\citep{joulin2016bag,joulin2016fasttext}, is used to categorize the language of the ASR transcripts; only samples where the audio language label matches the transcript language category are retained.

Furthermore, for specific audio types characterized by significant background noise—such as movies, TV shows, sports commentary, and esports commentary—we utilize MossFormer2~\citep{zhao2024mossformer2combiningtransformerrnnfree} for additional denoising.
Finally, for the training dataset, we strictly retain samples with DNSMOS $\ge 2.8$.

\subsection{Data Synthesis and Augmentation}

\textbf{Voice cloning data.} To equip the model with multi-speaker voice cloning capabilities, we construct a subset of data augmented with voice references. Specifically, for two-speaker and multi-speaker clips, we identify non-overlapping single-speaker segments extracted from the same raw recording. Based on the speaker identities derived from the diarization annotations, we map these single-speaker segments to the corresponding speaker reference slots within the prompt template. The chat templates used for both the voice cloning data and the common TTS data are provided in Appendix~\ref{app:chat_templates}.

\textbf{Synthesis data.} Accurately annotating speaker identities over long contexts in multi-speaker scenarios (3--5 speakers) poses a significant challenge, both for traditional speaker diarization models and for end-to-end systems like MOSS Transcribe Diarize that simultaneously generate speaker tags and ASR transcripts. Consequently, we filter our dataset, retaining only a small fraction of high-quality multi-speaker clips from real-world data. To enhance the model's speaker identity consistency over extended contexts, we augment the dataset with synthetic samples constructed by concatenating single-speaker clips. Specifically, single-speaker segments attributed to the same speaker by the diarization model are clustered into groups. These distinct speaker groups are then interleaved according to predefined rules to simulate the format of multi-speaker conversations. To ensure natural transitions in background acoustics when merging clips from diverse raw recordings, we strictly select segments with a DNSMOS score of $\ge 3.4$ and require all constituent clips within a single synthetic sample to share an identical sample rate.

\textbf{Text augmentation.} Certain typographical symbols are relatively scarce in real-world speech datasets, and Automatic Speech Recognition (ASR) models typically exhibit inherent biases toward generating a limited set of common punctuation marks. To mitigate these biases and improve the model's robustness to diverse text inputs, we apply rule-based text augmentation to enhance punctuation diversity in the transcripts of a selected subset of training samples. The specific substitution rules are detailed in Appendix~\ref{app:text_augmentation}.

\section{Method}

\begin{figure}[t]
  \centering
  \includegraphics[width=\linewidth]{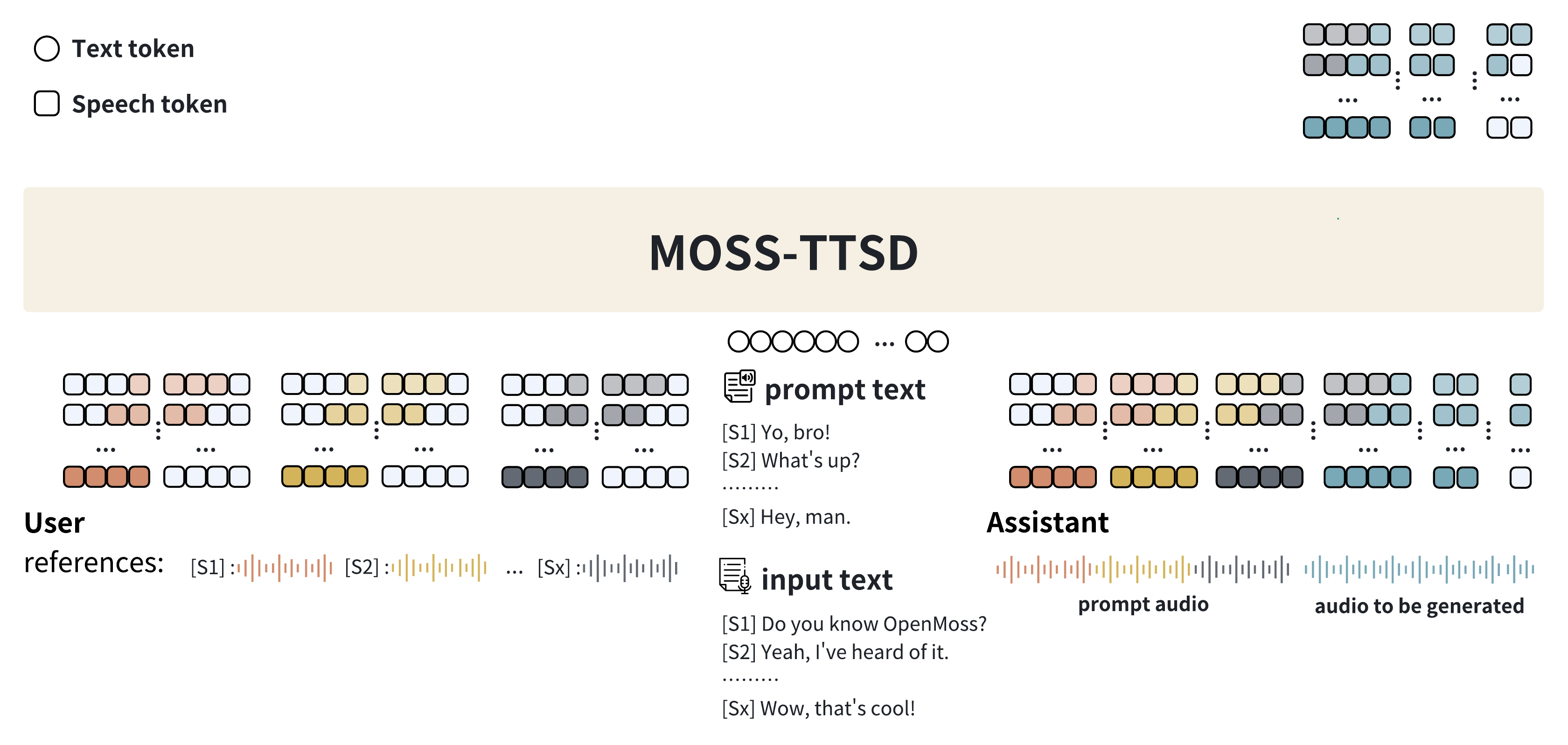}
  \caption{MOSS-TTSD inference for multi-speaker voice cloning. Given a text prompt with speaker tags, the model conditions on per-speaker reference audio and continues the spoken dialogue by generating discrete speech tokens, unifying reference-conditioned cloning with continuation-based cloning.}
  \label{fig:voice_clone_and_continuation}
\end{figure}

\subsection{Architecture}

MOSS-TTSD and MOSS-TTS~\citep{gong2026mossttstechnicalreport} share the same architecture and both adopt a fully discrete speech generation approach.
For sequence modeling, inspired by MusicGen, we use autoregressive generation with a multi-head delay pattern for speech tokens.

We use Qwen3-8B-base~\citep{Qwen3} as the LLM backbone and MOSS-Audio-Tokenizer~\citep{gong2026mossaudiotokenizerscalingaudiotokenizers} as the audio tokenizer.
Unlike MOSS-TTS, MOSS-TTSD models only the RVQ tokens from the first 16 layers of the audio tokenizer.
Thanks to the strong performance of MOSS-Audio-Tokenizer at a low bitrate of 2 kbps with a 12.5 Hz frame rate,
the LLM can robustly model audio sequences in long-context, multi-speaker settings, enabling training contexts
of up to 3600 seconds and up to five speakers.

\subsection{Curriculum Learning}
Starting from MOSS-TTS, MOSS-TTSD adopts a three-stage curriculum to shift from single-speaker synthesis to natural, high-quality, long-context multi-speaker dialogue generation.

\textbf{Stage 1.}
We perform continued pre-training from an intermediate MOSS-TTS checkpoint based on Qwen3-8B-base, pretrained on single-speaker TTS data with a sequence length of 32,768 tokens. In this stage, we include all single and two-speaker data with DNSMOS $\ge 2.8$ as well as voice cloning data, and extend the sequence length to 65,536 tokens. The goal is to adapt the model to longer contexts and to learn explicit control of speaker identity via speaker tags and reference audio, along with natural turn-taking in real dialogues.

\textbf{Stage 2.}
We restrict training to the subset of Stage 1 data with DNSMOS $\ge 3.4$ and true sample rate $\ge 24$ kHz, reduce the sampling ratio of single-speaker data, and lower the learning rate. This stage strengthens high-fidelity dialogue synthesis while preserving single-speaker capability.

\textbf{Stage 3.}
On top of all Stage 2 data, we add multi-speaker data from real raw recordings and a suitable proportion of synthetic data. This enables the model to generate high-quality speech for 1--5 speakers and improves turn-switching stability and speaker attribution accuracy with almost no degradation in speech naturalness.

\subsection{Inference}
Explicitly constructed voice cloning data teach the model to clone timbre from the reference slots defined in the chat templates. In addition, autoregressive TTS models naturally support voice cloning by continuation. We find that combining these two paradigms, as illustrated in Figure~\ref{fig:voice_clone_and_continuation}, substantially improves voice cloning performance; detailed results are reported in Appendix~\ref{app:voice_clone_and_continuation}. We define this setting as \texttt{voice\_clone\_and\_continuation} and use it as the default inference configuration.

\begin{figure}[t]
  \centering
  \includegraphics[width=\linewidth]{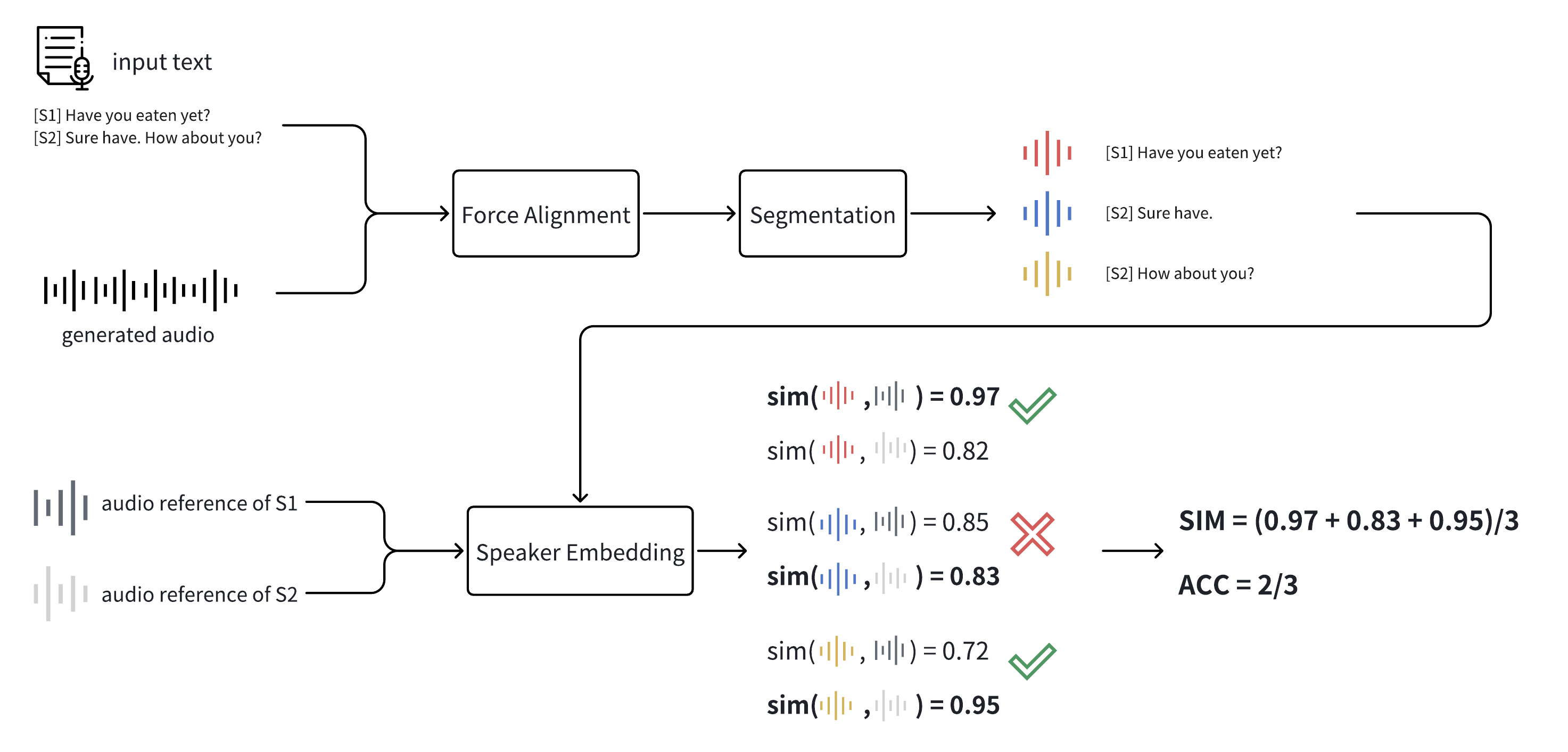}
  \caption{Overview of TTSD-eval. Given the input script with explicit speaker tags and the generated audio, TTSD-eval uses forced alignment to obtain word-level timestamps and segments the audio into utterance fragments. ACC and SIM are computed from speaker-embedding similarities between each fragment and the reference voices.}
  \label{fig:ttsd_eval_pipeline}
\end{figure}

\section{Evaluations}

\subsection{TTSD-eval}
\begin{table}[h]
  \centering
  \begin{threeparttable}
    \caption{TTSD-eval results. \textbf{ACC} denotes speaker attribution accuracy, \textbf{SIM} represents speaker voice similarity, and \textbf{WER} denotes Word Error Rate (lower is better). Best results are in \textbf{bold}.}
    \label{tab:dialogue_eval_full}
    \setlength{\tabcolsep}{3.5pt}
    \renewcommand{\arraystretch}{1.2}
    \begin{tabular}{lcccccc}
      \toprule
      \multirow{2}[0]{*}{\textbf{Model}} & \multicolumn{3}{c}{\textbf{ZH}} & \multicolumn{3}{c}{\textbf{EN}} \\
      \cmidrule(lr){2-4} \cmidrule(lr){5-7}
      & \textbf{ACC $\uparrow$} & \textbf{SIM $\uparrow$} & \textbf{WER $\downarrow$} & \textbf{ACC $\uparrow$} & \textbf{SIM $\uparrow$} & \textbf{WER $\downarrow$} \\
      \midrule
      \midrule

      \multicolumn{7}{l}{\textit{\textbf{Comparison with Open-Source Models}}} \\
      \midrule
      Higgs Audio V2 ~\citep{higgsaudio2025} & - & - & - & 0.9025 & 0.6860 & 21.31\% \\
      FireRedTTS-2 ~\citep{xie2025fireredtts} & 0.9022 & 0.7383 & 7.68\% & - & - & - \\
      VibeVoice 1.5B~\citep{peng2025vibevoice} & 0.8798 & 0.7415 & 8.18\% & 0.9353 & 0.6961 & 11.33\% \\
      VibeVoice 7B~\citep{peng2025vibevoice} & 0.9222 & 0.7590 & 5.70\% & 0.9554 & 0.7140 & \textbf{9.46\%} \\
      \textbf{MOSS-TTSD (ours)}& \textbf{0.9587} & \textbf{0.7949} & \textbf{4.85\%} & \textbf{0.9626} & \textbf{0.7326} & 9.88\% \\
      \midrule
      \midrule

      \multicolumn{7}{l}{\textit{\textbf{Comparison with Proprietary Models}}} \\
      \midrule
      Eleven V3 & 0.9653 & 0.6970 & \textbf{3.63\%} & 0.9498 & 0.6730 & \textbf{8.24\%} \\
      \textbf{MOSS-TTSD (elevenlabs\_voice)} & \textbf{0.9736} & \textbf{0.8165} & 3.91\% & \textbf{0.9565} & \textbf{0.7304} & 10.05\% \\
      \midrule
      gemini-2.5-pro-preview-tts & - & - & - & 0.9537 & 0.6786 & \textbf{8.59\%} \\
      gemini-2.5-flash-preview-tts & - & - & - & 0.9511 & 0.7194 & 8.71\% \\
      \textbf{MOSS-TTSD (gemini\_voice)} & - & - & - & \textbf{0.9655} & \textbf{0.7893} & 9.84\% \\
      \midrule
      Doubao\_Podcast & 0.9606 & 0.8034 & \textbf{4.72\%} & - & - & - \\
      \textbf{MOSS-TTSD (doubao\_voice)} & \textbf{0.9630} & \textbf{0.8226} & 5.71\% & - & - & - \\
      \bottomrule
    \end{tabular}
  \end{threeparttable}
\end{table}

Traditional evaluation of spoken dialogue generation typically reports cpWER~\citep{zhu2025zipvoicedialognonautoregressivespokendialogue,yu2025joyvoicelongcontextconditioninganthropomorphic} to indirectly assess speaker-attribution accuracy, and cpSIM to measure speaker similarity. However, both metrics are often constrained by the performance of speaker diarization models, and cpWER is further affected by the robustness of the ASR system. As the number of speakers increases from two to five or more, the error introduced by speaker diarization tends to grow substantially. To address these limitations, we propose TTSD-eval, an objective evaluation framework based on forced alignment. The overall TTSD-eval pipeline is illustrated in Figure~\ref{fig:ttsd_eval_pipeline}.

\textbf{Evaluation Metric.}
In TTSD-eval, we first employ MMS-FA (Meta's Massively Multilingual Speech Forced Alignment) to obtain word-level alignments between the input script and the generated audio. We then segment the audio into sentence fragments according to punctuation in the source text and assign speaker identities directly from the explicit speaker tags in the input script.

We adopt wespeaker-SimAMResNet100 as the speaker embedding model. For each generated fragment, we compute its similarity to the prompt audio of all candidate speakers and assign the speaker with the highest score as the prediction. By comparing these predicted labels with the ground-truth speaker tags in the input script, we calculate the \textbf{Speaker Attribution Accuracy (ACC)}. Furthermore, we define \textbf{Speaker Similarity (SIM)} as the similarity between each generated fragment and the prompt audio of its ground-truth speaker.

In addition, we report \textbf{Word Error Rate (WER)} to assess intelligibility. We use Whisper-large-v3 as the ASR system for both Chinese and English. To ensure a fair comparison, we strip all speaker tags and sound event tags, and further apply the text normalization procedure used in Seed-TTS-eva~\citep{seed_tts} to both the reference text and the ASR transcription before computing WER.

\textbf{Test Sets.}
Our test sets consist of both Chinese and English subsets, with 50 dialogue samples for each language. For both languages, we construct the corresponding subset using 20 pairs of speaker timbre references manually collected from the Internet and 30 reference pairs sampled from seed-tts-eval; all dialogue texts are generated using Gemini 2.5 Pro. Audio durations range from approximately 30 to 720 seconds, covering diverse scenarios from short dialogues to long-form multi-turn interactions, including podcasts, film and TV dubbing, sports and esports commentary, variety shows, animation dialogue, and crosstalk.

\textbf{Experimental Setup and Results.}
For open-source models, we directly follow the standard TTSD-eval pipeline and use the same test sets. Due to context-length and GPU-memory constraints, we set Higgs Audio V2's \texttt{generation\_chunk\_buffer\_size} to 6. Generation is performed chunk by chunk, and each new chunk is conditioned on the prompt and only the six most recently generated chunks. Under the same constraint, we implemented \texttt{generation\_chunk\_buffer\_size} for FireRedTTS2 following Higgs Audio V2 and also set it to 6.

For proprietary models, due to the constraints of voice cloning, we use speakers from each model's private voice library to ensure a fair comparison. MOSS-TTSD uses the same speakers as reference and performs zero-shot voice cloning for evaluation. We denote MOSS-TTSD with different reference voices as MOSS-TTSD (elevenlabs\_voice), MOSS-TTSD (gemini\_voice), and MOSS-TTSD (doubao\_voice); the corresponding voice information is provided in Appendix~\ref{app:proprietary_voices}. The texts used for evaluation are identical to those in the TTSD-eval test sets. Due to the Eleven V3 API's 5,000-character limit per request, we split the input into 10-turn chunks, generate audio per chunk, and concatenate them. Except for Eleven V3, all other models perform single-pass generation.

The results are summarized in Table~\ref{tab:dialogue_eval_full}, showing that MOSS-TTSD outperforms leading open-source and proprietary models in terms of speaker consistency and intelligibility in dialogue generation.

\begin{figure}[t]
  \centering
  \includegraphics[width=\linewidth]{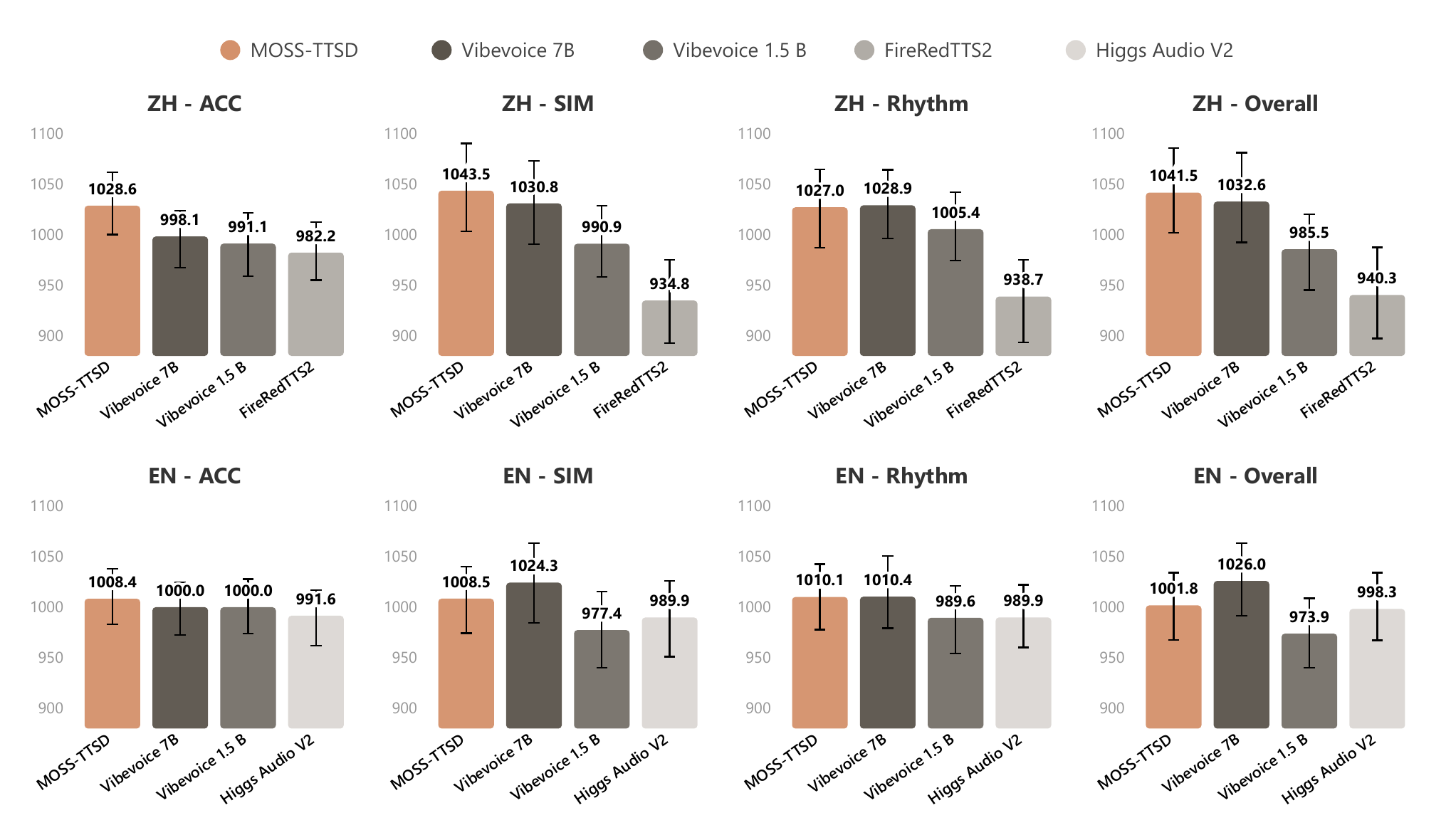}
  \caption{Elo ratings and confidence intervals of MOSS-TTSD and other open-source models on human-perceived speaker attribution accuracy (ACC), voice similarity (SIM), rhythm, and overall quality.}
  \label{fig:vs_open}
\end{figure}

\begin{figure}[t]
  \centering
  \includegraphics[width=\linewidth]{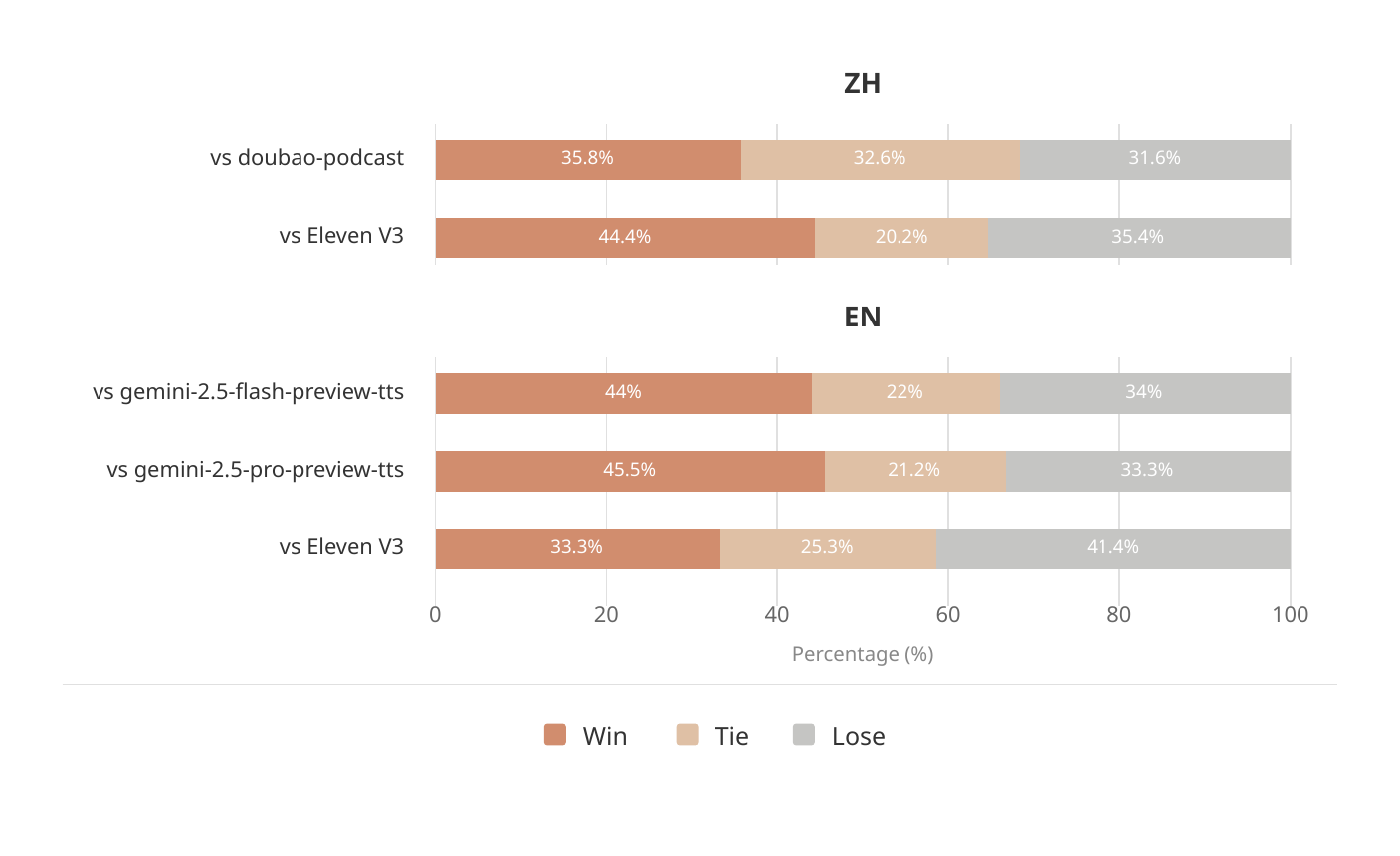}
  \caption{Subjective preference results between MOSS-TTSD and other proprietary models. Bars report win/tie/lose rates of MOSS-TTSD against proprietary baselines in Chinese (ZH) and English (EN), where annotators select the overall preferred sample.}
  \label{fig:vs_close}
\end{figure}

\subsection{Subjective Evaluation}

The subjective evaluation is conducted on the same test sets as TTSD-eval. To prevent annotators from losing focus when assessing long audio samples, we apply word-level alignment to segment each audio into clips no longer than 90 seconds for evaluation, and ensure that, for each rated audio pair, both samples correspond to the same text and speaker references.

For open-source models, annotators are asked to score each sample pair in terms of speaker attribution accuracy, voice similarity, rhythm, and overall quality. Following the methodology of the LMSYS Chatbot Arena~\citep{chiang2024chatbot}, we compute Elo ratings and confidence intervals for each dimension. The results are shown in Figure~\ref{fig:vs_open}.

For closed-source models, annotators are only asked to choose the overall preferred one in each pair, and we compute the win rate accordingly. The results are shown in Figure~\ref{fig:vs_close}.

Subjective evaluation results demonstrate that MOSS-TTSD surpasses leading open-source and proprietary models in terms of human-perceived quality. Furthermore, the strong correlation between human perception of speaker consistency and TTSD-eval scores validates the reliability of TTSD-eval as an effective objective metric.

\section{Conclusion}

In this work, we presented MOSS-TTSD, a spoken dialogue synthesis model for expressive and long-form multi-party conversations.
Built on a fully discrete paradigm with Qwen3-8B-base and MOSS-Audio-Tokenizer, MOSS-TTSD adopts a multi-head delay pattern and models only the first 16 RVQ layers to enable robust long-context generation.
Conditioned on dialogue scripts with explicit speaker tags and optional per-speaker references, it supports natural turn-taking, zero-shot multi-speaker voice cloning, and up to 5 speakers, scaling to 60 minutes of single-pass synthesis without stitching artifacts.
The model covers various mainstream languages and multiple long-form scenarios such as podcasts, commentary, dubbing, and entertainment content.
We further proposed TTSD-eval, a forced-alignment-based framework that measures speaker attribution accuracy (ACC) and speaker similarity (SIM) without relying on diarization tools, together with intelligibility via WER.
Both objective and subjective evaluations show that MOSS-TTSD consistently outperforms strong open-source and proprietary baselines for spoken dialogue generation.
We release our code and models to support future research and real-world deployment.

\section*{Contributors}
\noindent\textbf{Contributors:}\\
Yuqian Zhang\textsuperscript{*}, Donghua Yu, Zhengyuan Lin, Botian Jiang, Mingshu Chen, Yaozhou Jiang, Yiwei Zhao, Yiyang Zhang, Yucheng Yuan, Hanfu Chen, Kexin Huang, Jun Zhan, Cheng Chang, Zhaoye Fei, Shimin Li, Xiaogui Yang, Qinyuan Cheng.

\vspace{0.5em}
\noindent\textbf{Advisors:}\\
Xipeng Qiu\textsuperscript{$\dagger$}.

\vspace{0.5em}
\noindent\textbf{Affiliations:}\\
Shanghai Innovation Institute\\
MOSI Intelligence\\
Fudan University\\

\footnotetext[1]{* \href{mailto:yuqianzhang24@m.fudan.edu.cn}{\texttt{yuqianzhang24@m.fudan.edu.cn}}\hspace{2em}$\dagger$ Corresponding author: \href{mailto:xpqiu@fudan.edu.cn}{\texttt{xpqiu@fudan.edu.cn}}}

\clearpage
\bibliographystyle{unsrtnat}
\bibliography{main}

\clearpage
\appendix

\startcontents[app]
\begingroup
  \renewcommand{\contentsname}{Appendix Contents}
  \section*{\contentsname}
  \printcontents[app]{}{1}{}
\endgroup
\newpage

\section{Chat Templates of Training Data}
\label{app:chat_templates}

This appendix details the instruction-following chat templates used to construct training samples for both voice cloning and common TTS tasks. To ensure consistent conditioning across speakers and turns, both templates use explicit speaker tags in the text prompts. In addition, the voice cloning template includes per-speaker reference audio to guide timbre, whereas the common TTS template omits this field.

\begin{table}[htbp]
  \centering
  \caption{Chat templates for voice cloning data and common TTS data.}
  \label{tab:chat_templates}
  \renewcommand{\arraystretch}{1.3}
  \begin{tabular}{@{} p{0.5\textwidth} | p{0.45\textwidth} @{}}
    \toprule
    \textbf{\large Voice clone data} & \textbf{\large Common TTS data} \\
    \midrule
    \texttt{\textless|im\_start|\textgreater user} \newline
    \texttt{\textless user\_inst\textgreater} \newline
    \texttt{Reference(s):} \newline
    \quad\texttt{[S1]:} \textcolor{red}{\texttt{\textless audio}$_1$\texttt{\textgreater}} \newline
    \quad\texttt{[S2]:} \textcolor{red}{\texttt{\textless audio}$_2$\texttt{\textgreater}} \newline
    \quad\texttt{...} \newline
    \quad\texttt{[S}$_n$\texttt{]:} \textcolor{red}{\texttt{\textless audio}$_n$\texttt{\textgreater}} \newline
    \texttt{\{other fields: None\}} \newline
    \texttt{Text:} \newline
    \textcolor{blue}{\texttt{[S1]}\texttt{\{text}$_1$\texttt{\}}\texttt{[S2]}\texttt{\{text}$_2$\texttt{\}}...\texttt{[S}$_n$\texttt{]}\texttt{\{text}$_n$\texttt{\}}} \newline
    \texttt{\textless/user\_inst\textgreater\textless|im\_end|\textgreater} \newline
    \texttt{\textless|im\_start|\textgreater assistant} \newline
    \textcolor{blue}{\texttt{\textless generated\_audio\textgreater}} \newline
    \texttt{\textless|im\_end|\textgreater}
    &
    \texttt{\textless|im\_start|\textgreater user} \newline
    \texttt{\textless user\_inst\textgreater} \newline
    \texttt{Reference(s): None} \newline
    \texttt{\{other fields: None\}} \newline
    \texttt{Text:} \newline
    \textcolor{blue}{\texttt{[S1]}\texttt{\{text}$_1$\texttt{\}}\texttt{[S2]}\texttt{\{text}$_2$\texttt{\}}...\texttt{[S}$_n$\texttt{]}\texttt{\{text}$_n$\texttt{\}}} \newline
    \texttt{\textless/user\_inst\textgreater\textless|im\_end|\textgreater} \newline
    \texttt{\textless|im\_start|\textgreater assistant} \newline
    \textcolor{blue}{\texttt{\textless generated\_audio\textgreater}} \newline
    \texttt{\textless|im\_end|\textgreater} \\
    \bottomrule
  \end{tabular}
\end{table}

\section{Text Augmentation Rules}
\label{app:text_augmentation}

We apply rule-based punctuation augmentation to a subset of transcripts to improve robustness to diverse input styles. For each rule, the Original Token is replaced, with probability Prob., by a token randomly selected from the corresponding Replacement Set in Table~\ref{tab:text_augmentation}.

\begin{table}[htbp]
  \centering
  \caption{Rule-based text augmentation for punctuation diversity.}
  \label{tab:text_augmentation}
  \renewcommand{\arraystretch}{1.4}
  \begin{tabular}{@{} l c p{0.6\textwidth} @{}}
    \toprule
    \textbf{Original Token} & \textbf{Prob.} & \textbf{Replacement Set} \\
    \midrule
    Comma ([\texttt{,}] or [\cjk{\texttt{，}}]) & 10\% &
    [\cjk{\texttt{、}}] \quad [\texttt{;}] \quad [\cjk{\texttt{；}}] \quad [\cjk{\texttt{：}}] \quad [\texttt{:}] \quad [\texttt{-}] \quad [\texttt{\_}] \quad [\texttt{---}] \quad [\texttt{--}] \quad [\texttt{―}] \quad [\texttt{\textbackslash n}] \quad [\textit{space}] \\
    \addlinespace
    Period ([\texttt{.}] or [\cjk{\texttt{。}}]) & 5\% &
    [\texttt{\dots\dots}] \quad [\texttt{......}] \quad [\texttt{\dots}] \quad [\texttt{...}] \quad [\cjk{\texttt{～}}] \quad [\texttt{\textasciitilde}] \quad [\texttt{\textbackslash n}] \quad [\textit{space}] \\
    \bottomrule
  \end{tabular}
\end{table}

\section{Voice Clone and Continuation}
\label{app:voice_clone_and_continuation}

Table~\ref{tab:seed_tts_eval} reports Seed-TTS-eval results and Table~\ref{tab:voice_clone_and_continuation} reports TTSD-eval results under the same three inference configurations: \texttt{voice\_clone} (reference-conditioned cloning only), \texttt{continuation} (pure autoregressive continuation), and \texttt{voice\_clone\_and\_continuation} (the default setting that combines both).

From the TTSD-eval and Seed-TTS-eval results, the combined \texttt{voice\_clone\_and\_continuation} setting achieves the best speaker similarity and obtains comparable results on the other two metrics.

\begin{table}[h]
  \centering
  \begin{threeparttable}
    \caption{Seed-TTS-eval results. \textbf{SIM} denotes speaker voice similarity (higher is better). Best results are in \textbf{bold}.}
    \label{tab:seed_tts_eval}
    \setlength{\tabcolsep}{6pt}
    \renewcommand{\arraystretch}{1.2}
    \begin{tabular}{lccc}
      \toprule
      \textbf{Model} & \textbf{EN SIM $\uparrow$} & \textbf{ZH SIM $\uparrow$} & \textbf{ZH (hard case) SIM $\uparrow$} \\
      \midrule
      \texttt{MOSS-TTSD(voice\_clone)} & 0.6075 & 0.7160 & 0.6964 \\
      \texttt{MOSS-TTSD(continuation)} & 0.6579 & 0.7513 & 0.7248 \\
      \texttt{MOSS-TTSD(voice\_clone\_and\_continuation)} & \textbf{0.6828} & \textbf{0.7590} & \textbf{0.7401} \\
      \bottomrule
    \end{tabular}
  \end{threeparttable}
\end{table}

\begin{table}[h]
  \centering
  \begin{threeparttable}
    \caption{TTSD-eval results for different voice cloning paradigms. \textbf{ACC} denotes speaker attribution accuracy, \textbf{SIM} represents speaker voice similarity, and \textbf{WER} denotes Word Error Rate (lower is better). Best results are in \textbf{bold}.}
    \label{tab:voice_clone_and_continuation}
    \setlength{\tabcolsep}{3.5pt}
    \renewcommand{\arraystretch}{1.2}
    \begin{tabular}{lcccccc}
      \toprule
      \multirow{2}[0]{*}{\textbf{Model}} & \multicolumn{3}{c}{\textbf{ZH}} & \multicolumn{3}{c}{\textbf{EN}} \\
      \cmidrule(lr){2-4} \cmidrule(lr){5-7}
      & \textbf{ACC $\uparrow$} & \textbf{SIM $\uparrow$} & \textbf{WER $\downarrow$} & \textbf{ACC $\uparrow$} & \textbf{SIM $\uparrow$} & \textbf{WER $\downarrow$} \\
      \midrule
      \texttt{MOSS-TTSD(voice\_clone)} & 0.9387 & 0.7852 & 6.07\% & \textbf{0.9680} & 0.7228 & 9.39\% \\
      \texttt{MOSS-TTSD(continuation)} & 0.9254 & 0.7740 & 5.46\% & 0.9209 & 0.7107 & \textbf{9.07\%} \\
      \texttt{MOSS-TTSD(voice\_clone\_and\_continuation)} & \textbf{0.9587} & \textbf{0.7949} & \textbf{4.85\%} & 0.9626 & \textbf{0.7326} & 9.88\% \\
      \bottomrule
    \end{tabular}
  \end{threeparttable}
\end{table}

\section{Voice References used for Evaluation}
\label{app:proprietary_voices}

Table~\ref{tab:proprietary_voice_refs} summarizes the proprietary voice references used in our evaluations. For Doubao, the official voice library does not provide speaker prompts for \textit{Doubao\_Podcast}; therefore, we use the official API to generate two corresponding speaker prompts as the references in our evaluation.

\begin{table}[h]
  \centering
  \caption{Proprietary voice references used for evaluation.}
  \label{tab:proprietary_voice_refs}
  \setlength{\tabcolsep}{4pt}
  \renewcommand{\arraystretch}{1.05}
  \small
  \begin{threeparttable}
    \begin{tabular}{l|l}
      \toprule
      \textbf{Reference} & \textbf{Voice} \\
      \midrule
      \midrule

      \multicolumn{2}{l}{\textbf{ZH}} \\
      \midrule
      elevenlabs\_voice &
      \makecell[l]{speaker1: Siqi Liu -- Calm, Warm and Gentle (Voice ID: W8lBaQb9YIoddhxfQNLP)\\
                   speaker2: Susan -- Clear and Calm Storyteller (Voice ID: kAIqZ7fZv234ClKXwzDx)} \\
      \midrule
      doubao\_voice &
      \makecell[l]{speaker1: zh\_male\_dayixiansheng\_v2\_saturn\_bigtts\\
                   speaker2: zh\_female\_mizaitongxue\_v2\_saturn\_bigtts} \\
      \midrule
      \midrule
      \multicolumn{2}{l}{\textbf{EN}} \\
      \midrule
      
      elevenlabs\_voice &
      \makecell[l]{speaker1: Henry -- Deep, Professional, and Soothing (Voice ID: pVnrL6sighQX7hVz89cp)\\
                   speaker2: Ava -- Eager, Helpful and Understanding (Voice ID: gJx1vCzNCD1EQHT212Ls)} \\
      \midrule
      gemini\_voice &
      \makecell[l]{speaker1: Puck\\
                   speaker2: Zephyr} \\
      \bottomrule
    \end{tabular}
  \end{threeparttable}
\end{table}

\end{document}